\documentclass[conference]{IEEEtran}
\IEEEoverridecommandlockouts

\usepackage{cite}
\usepackage{amsmath,amssymb,amsfonts}
\usepackage{algorithmic}
\usepackage{graphicx}
\graphicspath{{figures/}{./}}
\usepackage{textcomp}
\usepackage{xcolor}
\usepackage{url}
\usepackage{booktabs}
\usepackage{array}
\usepackage{mathtools}

\def\BibTeX{{\rm B\kern-.05em{\sc i\kern-.025em b}\kern-.08em
    T\kern-.1667em\lower.7ex\hbox{E}\kern-.125emX}}

\newcommand{\vect}[1]{\mathbf{#1}}
\newcommand{\uvect}[1]{\hat{\mathbf{#1}}}
\newcommand{\norm}[1]{\left\lVert #1 \right\rVert}
\newcommand{\Transpose}{\mathsf{T}}
\newcommand{\inlineeqnum}[2]{\refstepcounter{equation}\label{#1}#2~(\theequation)}
\DeclareMathOperator*{\argmax}{arg\,max}

\DeclareMathOperator{\diag}{diag}

\begin{document}

\title{Performance Evaluation of Mobile Target Sensing and Tracking Across Multiple Sensing Modes}

\author{
\IEEEauthorblockN{1\textsuperscript{st} Necati Kagan Erkek}
\IEEEauthorblockA{
\textit{InterDigital, Inc.} \\
\textit{King's College London} \\
London WC2R2LS, United Kingdom  \\
necati.erkek@kcl.ac.uk
}

\and

\IEEEauthorblockN{\hspace{0.5cm}2\textsuperscript{nd} Marco Di Renzo}
\IEEEauthorblockA{
\textit{\hspace{0.5cm}King's College London}\\
\hspace{0.5cm}London WC2R2LS, United Kingdom \\
\textit{\hspace{0.5cm}CNRS and CentraleSup\'elec, IETR} \\ 
\hspace{0.5cm} Avenue de la Boulaie, 35576 Cesson-S\'evign\'e, France
} 
\and 
\hspace{0.5cm}
\IEEEauthorblockN{3\textsuperscript{rd} Arman Shojaeifard}
\IEEEauthorblockA{\hspace{0.5cm}
\textit{InterDigital, Inc.} \\
\hspace{0.5cm} London EC3A3DH, UK \\
}
\and
\IEEEauthorblockN{4\textsuperscript{th} Kunjan Shah}
\IEEEauthorblockA{
\textit{InterDigital, Inc.} \\
London EC3A3DH, UK\\
}
\and
\IEEEauthorblockN{5\textsuperscript{th} Mohammad Heggo}
\IEEEauthorblockA{
\textit{InterDigital, Inc.} \\
London EC3A3DH, UK \\
}
\and
\IEEEauthorblockN{6\textsuperscript{th} Yasser Mestrah}
\IEEEauthorblockA{
\textit{InterDigital, Inc.} \\
London EC3A3DH, UK \\
}
}

\maketitle

\begin{abstract}
Moving extended-target sensing and tracking are studied in a beamformed PRS/OFDM ISAC framework. The target is modeled as a rigid cluster of scatterers whose Tx--target--Rx echoes coherently combine in a MIMO scattering channel. Delay, azimuth/elevation, and Doppler are estimated using PRS matched filtering with delay centroiding, single-cluster 2D-MUSIC, and FFT-based Doppler estimation, then used for independent localization and EKF-based tracking. The framework is evaluated in monostatic and bistatic modes under target-directed, receiver-directed, and omnidirectional beamforming. Simulation results show that EKF tracking reduces the aggregate 3D RMSE from $1.61$ m to $1.38$ m relative to independent localization. Target-directed beamforming substantially improves tracking accuracy, achieving $1.38$ m RMSE compared with $5.91$ m and $8.38$ m for omnidirectional and receiver-directed beamforming, respectively.  The results highlight the importance of target-aware beam management, delay-angle-Doppler fusion, and sensing geometry for mobile-target tracking in ISAC networks.
\end{abstract}


\section{Introduction}
\IEEEPARstart{I}{ntegrated} sensing and communication (ISAC) is emerging as a key beyond-5G/6G concept by reusing spectrum, hardware, waveforms, and infrastructure for both connectivity and environmental awareness \cite{ref_liu_tcom_2020,ref_liu_jsac_2022,ref_zhang_survey_2022}.  By exploiting radio-frequency reflections from the environment, the same radio access infrastructure can support communication together with sensing functions such as detection, localization, and tracking of passive or mobile objects. In perceptive mobile networks, TRPs, UEs, and network-side sensing functions can cooperate through co-located or distributed sensing modes, including TRP--UE bistatic sensing and hybrid TRP--UE/TRP--TRP localization~\cite{ref_erkek_ojcoms_2026,ref_erkek_vtc2026_hybrid}. In parallel, 3GPP has standardized 5G NR positioning, including PRS, with Release-18 further improving accuracy and device support \cite{ref_cha_rel18,ref_italiano_tutorial}. In FR2, large bandwidths and arrays enable fine delay and angular resolution, while directional beams make beam selection critical for echo strength and sensing geometry \cite{ref_italiano_tutorial,ref_shahmansoori_twc_2018,ref_wymeersch_globecom_2018}. 

However, sensing becomes more challenging when the target is extended rather than point-like, because multiple scatterers contribute to the echo \cite{ref_wang_extended_target,ref_wang_3d_extended_target,ref_luo_moving_target}. Extended-target formulations are needed to describe how the measured echo is formed by coherent superposition of scatterer-induced reflections carried by the same moving object, rather than by a single ideal point response \cite{ref_wang_extended_target,ref_wang_3d_extended_target,ref_pedraza_extended_tracking}. Such modeling is relevant because delay, azimuth/elevation, and Doppler depend jointly on transmitter--target and target--receiver geometries, affecting measurements and tracking performance. Existing ISAC tracking studies often consider point-like targets, isolated monostatic or bistatic geometries, or idealized beam alignment, leaving unclear how PRS-based delay-angle-Doppler extraction performs for extended mobile targets across different sensing modes and beamforming strategies.

The main contributions are fourfold. First, a beamformed PRS/OFDM signal model and MIMO scattering channel are developed for a moving extended target represented by a rigid cluster of scatterers; the model explains how the sensing target arises from superposition of scatterer-induced echoes. Second, delay, azimuth/elevation, and Doppler measurements are derived from PRS matched filtering with delay centroiding, single-cluster 2D-MUSIC, and FFT-based Doppler estimation. Third, two tracking assumptions are formulated: independent time-stamp tracking without temporal filtering and extended Kalman filter (EKF)-filtered tracking based on path length, azimuth, elevation, and Doppler measurements. Fourth, tracking performance is compared under three beamforming mechanisms: target or receiver directed beams, and omnidirectional.

\section{Signal, Noise, and Channel Model}

\begin{figure*}[!t]
    \centering
    \includegraphics[width=1\linewidth]{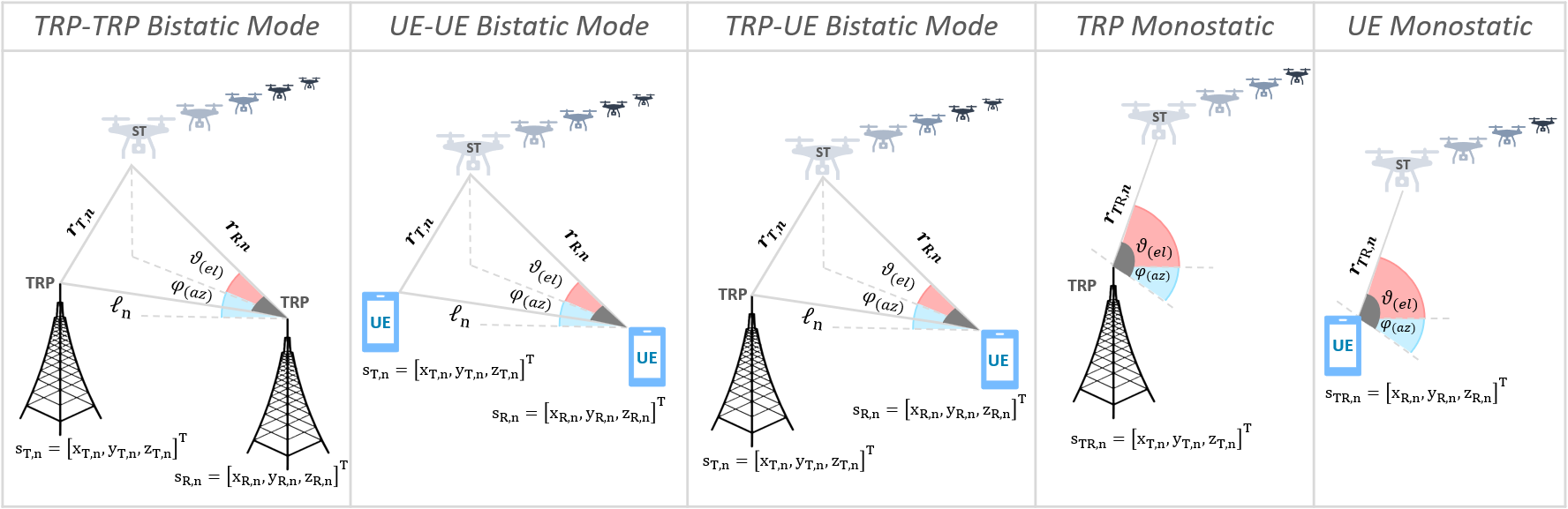}
    \caption{Illustration of different sensing modes for mobile target tracking.}
    \label{SensingModes}
\end{figure*}

\subsection{Signal Model: Beamformed PRS/OFDM Waveform}

Let $X[k,m]$ denote PRS symbol placed on subcarrier $k$ and OFDM symbol $m$, where $k \in \mathcal{K}_{\mathrm{PRS}}$ and $m=0,\ldots,M-1$. The transmitted scalar complex-baseband waveform is:
\begin{equation}
s(t) = \sum_{m=0}^{M-1} \sum_{k=0}^{K_{PRS}}X_{[k,m]} \, g(t-mT_{\mathrm{sym}})e^{j2\pi k\Delta f (t-mT_{\mathrm{sym}})}
\label{eq:ofdm_signal}
\end{equation}

where $\Delta f$ is the subcarrier spacing, $T_{\mathrm{sym}}$ is the OFDM symbol duration including cyclic prefix, and $g(t)$ represents the corresponding pulse shape. If the PRS occupies $N_{\mathrm{RB}}$ resource blocks, the occupied bandwidth is $B = 12 N_{\mathrm{RB}} \Delta f$. The associated delay resolution of the matched-filtered waveform is of order $1/B$. Assume an $N_{\mathrm{T}}$-element transmit array and an $N_{\mathrm{R}}$-element receive array. With transmit beamforming vector $\vect{w}_{\mathrm{T}} \in \mathbb{C}^{N_{\mathrm{T}}}$ satisfying $\norm{\vect{w}_{\mathrm{T}}}_2=1$, the radiated vector signal is $\vect{x}(t)=\vect{w}_{\mathrm{T}}s(t)$. The receive array signal before combining is denoted by $\vect{y}(t)$, and an optional receive beamforming vector $\vect{w}_{\mathrm{R}} \in \mathbb{C}^{N_{\mathrm{R}}}$ produces the scalar observation $y(t)=\vect{w}_{\mathrm{R}}^{\mathsf{H}}\vect{y}(t),\ \norm{\vect{w}_{\mathrm{R}}}_2=1$. The full array output $\vect{y}(t)$ is used for angle estimation, while the beamformed scalar output $y(t)$ is used for delay and Doppler extraction. The same model is evaluated under three beamforming mechanisms for performance comparison: ST-directed beamforming, Rx-directed beamforming, and omnidirectional operation, which are represented by the corresponding choices of $\vect{w}_{\mathrm{T}}$ and $\vect{w}_{\mathrm{R}}$.

\subsection{Sensing Target Model: Cluster of Scatterers}
A mobile sensing target is modeled as an extended object composed of $Q$ scattering centers. Let $\vect{p}[n] \in \mathbb{R}^3$ denote the target-center position at slow-time index $n$. A general rigid-body model can be written as:
\begin{equation}
\vect{s}_q[n] = \vect{p}[n] + \vect{R}[n]\boldsymbol{\delta}_q,
\qquad q=1,\ldots,Q,
\label{eq:rigid_body_scatterers}
\end{equation}
where $\boldsymbol{\delta}_q$ is the $q$th scatterer offset in the target body frame and $\vect{R}[n]$ is the target orientation matrix. For rigid translation without rotation, $\vect{R}[n]=\vect{I}_3$ and therefore $\vect{s}_q[n]=\vect{p}[n]+\boldsymbol{\delta}_q$. Hence all scatterers share the same translational velocity $\vect{v}[n]=\dot{\vect{p}}[n]$. This discrete-scatterer representation is standard for extended-target sensing because a physically large object can generate multiple reflections whose delays, phases, and angles evolve jointly with the rigid-body motion \cite{ref_wang_extended_target,ref_wang_3d_extended_target,ref_luo_moving_target,ref_pedraza_extended_tracking}. Let $\vect{p}_{\mathrm{T}}$ and $\vect{p}_{\mathrm{R}}$ denote the transmitter and receiver positions, respectively. For the $q$th scatterer, define the transmitter-to-scatterer and scatterer-to-receiver distances:
\begin{align}
r_{\mathrm{T},q}[n] = \norm{\vect{s}_q[n]-\vect{p}_{\mathrm{T}}}, \qquad
r_{\mathrm{R},q}[n] = \norm{\vect{s}_q[n]-\vect{p}_{\mathrm{R}}}
\label{eq:rTq}
\end{align}
The propagation delay is $\tau_q[n]=(r_{\mathrm{T},q}[n]+r_{\mathrm{R},q}[n])/c$, where $c$ is the speed of light.

\subsection{MIMO Scattering Channel and Noise Model}
Let $\Omega_{\mathrm{T},q}[n]$ and $\Omega_{\mathrm{R},q}[n]$ denote the departure and arrival angles of the $q$th scatterer, and let $\vect{a}_{\mathrm{T}}(\Omega)$ and $\vect{a}_{\mathrm{R}}(\Omega)$ be the corresponding steering vectors. The channel model adopts a narrowband steering approximation, so the frequency dependence of the array steering vectors across the PRS band is neglected. For the specified simulation parameters ($B=500$ MHz at a $35$ GHz carrier), the fractional bandwidth is about $1.4\%$; possible beam-squint and other wideband-array effects are therefore outside the present model. The $q$th scatterer coefficient is:
\begin{equation}
\beta_q[n] = \frac{\rho_q[n]}{r_{\mathrm{T},q}[n]r_{\mathrm{R},q}[n]}
\exp\!\big[-j\frac{2\pi}{\lambda}
\big(r_{\mathrm{T},q}[n]+r_{\mathrm{R},q}[n]\big)\big]
\label{eq:scatterer_gain}
\end{equation}
where $\rho_q[n]$ is the complex reflectivity and $\lambda$ is the carrier wavelength. The Doppler shift follows from the time derivative of the total propagation length:
\begin{equation}
f_{\mathrm{D},q}[n]
= -\frac{1}{\lambda}\frac{d}{dt}
\big(r_{\mathrm{T},q}[n]+r_{\mathrm{R},q}[n]\big)
\label{eq:doppler_derivative}
\end{equation}
Using \eqref{eq:rTq}, define the unit vectors:
\begin{equation}
\uvect{u}_{\mathrm{T},q}[n]
= \frac{\vect{s}_q[n]-\vect{p}_{\mathrm{T}}}{r_{\mathrm{T},q}[n]},
\qquad
\uvect{u}_{\mathrm{R},q}[n]
= \frac{\vect{s}_q[n]-\vect{p}_{\mathrm{R}}}{r_{\mathrm{R},q}[n]}
\label{eq:unit_vectors}
\end{equation}
Since all scatterers share the same velocity $\vect{v}[n]$, the Doppler:
\begin{equation}
f_{\mathrm{D},q}[n]
= -\frac{1}{\lambda}\vect{v}^{\Transpose}[n]
\big(\uvect{u}_{\mathrm{T},q}[n]
+ \uvect{u}_{\mathrm{R},q}[n]\big) 
\label{eq:doppler_scatterer}
\end{equation}
The equivalent baseband channel impulse response is then
\begin{equation}
\begin{aligned}
\vect{H}(t,\tau;n)
&= \vect{H}_{\mathrm{ref}}(t,\tau;n) \\
&\quad + \sum_{q=1}^{Q} \beta_q[n]
\vect{a}_{\mathrm{R}}\!\big(\Omega_{\mathrm{R},q}[n]\big)
\vect{a}_{\mathrm{T}}^{\mathsf{H}}\!\big(\Omega_{\mathrm{T},q}[n]\big) \\
&\quad \times e^{j2\pi f_{\mathrm{D},q}[n] t}
\delta\big(\tau-\tau_q[n]\big),
\end{aligned}
\label{eq:mimo_channel}
\end{equation} 
where $\vect{H}_{\mathrm{ref}}(t,\tau;n)$ is a reference path that can be used for synchronization or delay calibration.
Ideal time--frequency synchronization and delay calibration are assumed for bistatic measurements; residual clock and carrier-frequency offsets are excluded. The delay $\hat{\tau}[n]$ is the absolute Tx--target--Rx delay: reference-relative measurements require restoration of the known reference-path delay. A residual timing offset $\Delta\tau$ would bias the total path length by $c\Delta\tau$.
The received signal:
\begin{equation}
\vect{y}(t;n) = \int \vect{H}(t,\tau;n)\vect{x}(t-\tau)\,d\tau + \vect{n}(t;n)
\label{eq:rx_array_signal}
\end{equation}
with additive array noise \inlineeqnum{eq:noise_model}{$\vect{n}(t;n)\sim\mathcal{CN}\big(\vect{0},\sigma_n^2\vect{I}_{N_{\mathrm{R}}}\big)$}. Substituting $\vect{x}(t)=\vect{w}_{\mathrm{T}}s(t)$ into \eqref{eq:rx_array_signal}, the target-induced array response:
\begin{equation}
\begin{aligned}
\vect{y}_{\mathrm{tar}}(t;n)
&= \sum_{q=1}^{Q}
\alpha_q[n]\,\vect{a}_{\mathrm{R}}\!\big(\Omega_{\mathrm{R},q}[n]\big)
 s\big(t-\tau_q[n]\big)e^{j2\pi f_{\mathrm{D},q}[n] t}
\end{aligned}
\label{eq:array_target_echo}
\end{equation}
where \inlineeqnum{eq:alpha_q}{$\alpha_q[n]=\beta_q[n]\,\vect{a}_{\mathrm{T}}^{\mathsf{H}}\!\big(\Omega_{\mathrm{T},q}[n]\big)\vect{w}_{\mathrm{T}}$}. After receive combining, the scalar echo is:
\begin{equation}
\begin{aligned}
y_{\mathrm{tar_{(t;n)}}}
&= \sum_{q=1}^{Q} \gamma_{q_{[n]}}
 s\big(t-\tau_q[n]\big)e^{j2\pi f_{\mathrm{D},q}[n] t} + \eta_{(t;n)}
\end{aligned}
\label{eq:scalar_target_echo}
\end{equation}
with effective coefficient \inlineeqnum{eq:gamma_q}{$\gamma_q[n]=\vect{w}_{\mathrm{R}}^{\mathsf{H}}\vect{a}_{\mathrm{R}}\!\big(\Omega_{\mathrm{R},q}[n]\big)\alpha_q[n]$} and scalar noise $\eta(t;n)=\vect{w}_{\mathrm{R}}^{\mathsf{H}}\vect{n}(t;n)$. Equations \eqref{eq:mimo_channel}--\eqref{eq:gamma_q} show that the sensing target is not inserted as a single abstract reflection coefficient; instead, it is created by the coherent superposition of the scatterer-induced Tx--scatterer--Rx paths.

\section{Delay, Doppler, and Angular Measurements}

\subsection{Delay Based Measurement from PRS Matched Filtering}
Delay sensing starts from the echo in \eqref{eq:scalar_target_echo}. Correlating it with PRS waveform gives \inlineeqnum{eq:matched_filter_output}{$\chi(\tau;n)=\int y(t;n)s^*(t-\tau)\,dt$}. Substitution yields the PRS ambiguity-function form:
\begin{equation}
\chi(\tau;n)
= \sum_{q=1}^{Q} \gamma_q[n]
\mathcal{A}_s\!\big(\tau-\tau_q[n],f_{\mathrm{D},q}[n]\big)
+ \widetilde{\eta}(\tau;n),
\label{eq:ambiguity_sum}
\end{equation}

\begin{equation}
\mathcal{A}_s(\tau,\nu)
= \int s(t)s^*(t-\tau)e^{j2\pi \nu t}\,dt
\label{eq:ambiguity_function}
\end{equation}
Since the Doppler phase is small over one PRS burst, the zero-Doppler slice gives the delay profile:
\begin{equation}
\chi(\tau;n)
\approx \sum_{q=1}^{Q} \gamma_q[n]R_s\big(\tau-\tau_q[n]\big)
+ \widetilde{\eta}(\tau;n),
\label{eq:delay_approximation}
\end{equation}
where $R_s(\tau)=\mathcal{A}_s(\tau,0)$. The power delay profile is \inlineeqnum{eq:pdp_definition}{$P_{\mathrm{PDP}}(\tau;n)=\left|\chi(\tau;n)\right|^2$}. For the extended target, the delay cluster in window $\mathcal{D}_n$ is represented by the centroid:
\begin{equation}
\bar{\tau}[n]
= \frac{\int_{\mathcal{D}_n} \tau P_{\mathrm{PDP}}(\tau;n)\,d\tau}
{\int_{\mathcal{D}_n} P_{\mathrm{PDP}}(\tau;n)\,d\tau}
\label{eq:delay_centroid_continuous}
\end{equation}
With the discrete scatterer model and negligible cross-terms, this becomes the power-weighted form:
\begin{equation}
\bar{\tau}[n] = \sum_{q=1}^{Q} \omega_q[n]\tau_q[n],
\qquad
\omega_q[n] = \frac{|\gamma_q[n]|^2}{\sum_{i=1}^{Q}|\gamma_i[n]|^2}
\label{eq:delay_centroid_discrete}
\end{equation}
The noisy delay and path-length measurements can be written as \inlineeqnum{eq:delay_measurement}{$\hat{\tau}[n]=\bar{\tau}[n]+\eta_{\tau}[n]$} and \inlineeqnum{eq:path_length_measurement}{$\hat{R}_{\Sigma}[n]=c\hat{\tau}[n]$}.

\subsection{Angular Measurement via 2D-MUSIC}

Angle estimation uses the receive-array snapshots before scalar combining. The scatterer cluster forms the response:
\begin{equation}
\bar{\vect{a}}_{\mathrm{R}}[n]
= \sum_{q=1}^{Q} \alpha_q[n]
\vect{a}_{\mathrm{R}}\!\big(\phi_q[n],\theta_q[n]\big),
\label{eq:effective_steering}
\end{equation}
Here $(\phi_q[n],\theta_q[n])$ are receiver-side scatterer angles. Under a narrow-angular-spread approximation, the cluster is treated as one dominant spatial source. The sample covariance provides the noise subspace for a standard 2D-MUSIC angular-grid search \cite{ref_schmidt_music}. The selected azimuth/elevation pair represents the dominant cluster direction, not individual scatterers.

\subsection{Doppler Measurement via Spectral Analysis}
After delay gating and receive combining, the slow-time samples collected with $T_{\mathrm{r}}=1/f_{\mathrm{PRF}}$ are modeled as:
\begin{equation}
z[\ell;n]=\sum_{q=1}^{Q}\gamma_q[n]
e^{j2\pi f_{\mathrm{D},q}[n]\ell T_{\mathrm{r}}}+\upsilon_{[\ell;n]},
\ \ell=0,.,L_{\mathrm{s}}-1 
\label{eq:slow_time_signal}
\end{equation}
An $N_{\mathrm{FFT}}$-point FFT of the windowed sequence $\varpi[\ell]z[\ell;n]$, followed by frequency centering, produces $Z[k;n]$ on the grid $f_k=(k-N_{\mathrm{FFT}}/2)f_{\mathrm{PRF}}/N_{\mathrm{FFT}}$, $k=0,\ldots,N_{\mathrm{FFT}}-1$.
The Doppler estimate is the dominant peak $\hat{f}_{\mathrm{D}}[n]=f_{k^{\star}}$, where $k^{\star}=\argmax_k|Z[k;n]|^2$.
For target, define the path-gradient vector and derive Doppler:
\begin{equation}
\vect{g}(\vect{p})=
\frac{\vect{p}-\vect{p}_{\mathrm{T}}}{\norm{\vect{p}-\vect{p}_{\mathrm{T}}}}
+\frac{\vect{p}-\vect{p}_{\mathrm{R}}}{\norm{\vect{p}-\vect{p}_{\mathrm{R}}}}
\label{eq:geometry_vector}
\end{equation}
\begin{equation}
f_{\mathrm{D}}[n]=-
\lambda^{-1}\vect{v}^{\Transpose}[n]\vect{g}\big(\vect{p}[n]\big)
\label{eq:ideal_path_doppler}
\end{equation}
Thus one Doppler value observes only the velocity projection along $\vect{g}$. For the scalar per-snapshot velocity descriptor, imposing the vector \inlineeqnum{eq:x_axis_motion}{$\vect{v}[n]=\big[v_x[n],0,0\big]^{\Transpose}$} yields with \inlineeqnum{eq:vx_from_doppler}{$\hat{v}_x[n]=-\lambda\hat{f}_{\mathrm{D}}[n]/g_x\big(\hat{\vect{p}}[n]\big)$} when $g_x\big(\hat{\vect{p}}[n]\big)\neq0$.
\vspace{4pt}
\section{Tracking Theory}
\vspace{4pt}
Tracking input is
$
\vect{z}[n] =
\begin{bmatrix}
\hat{R}_{\Sigma}[n] & \hat{\phi}[n] & \hat{\theta}[n] & \hat{f}_{\mathrm{D}}[n]
\end{bmatrix}^{\Transpose}
$
The first three entries constrain position, while Doppler constrains velocity through the Tx--target--Rx geometry.

\subsection{Independent Time-Stamp Tracking}
Each snapshot is first localized without temporal smoothing. The receive-side bearing from the measured azimuth and elevation pair can be presented as:

\begin{equation}
\hat{\vect{u}}_{[n]}=
\begin{bmatrix}
\cos\hat{\theta}_{[n]}\cos\hat{\phi}_{[n]} &
\cos\hat{\theta}_{[n]}\sin\hat{\phi}_{[n]} &
\sin\hat{\theta}_{[n]}
\end{bmatrix}^{\Transpose}
\label{eq:bearing_unit_vector}
\end{equation}
The target is placed on the bearing line $\vect{p}=\vect{p}_{\mathrm{R}}+r\hat{\vect{u}}[n]$, $r\ge0$, and constrained by the measured path length is: 

\begin{equation}
\norm{\vect{p}-\vect{p}_{\mathrm{T}}}+\norm{\vect{p}-\vect{p}_{\mathrm{R}}}=\hat{R}_{\Sigma}[n]
\label{eq:path_length_constraint}
\end{equation}
Let $\vect{d}=\vect{p}_{\mathrm{R}}-\vect{p}_{\mathrm{T}}$, $D=\norm{\vect{d}}^2$, and $a[n]=\hat{\vect{u}}^{\Transpose}[n]\vect{d}$. Substitution gives $\norm{\vect{d}+r\hat{\vect{u}}[n]}=\hat{R}_{\Sigma}[n]-r$; after squaring, the line distance, position estimate and the Doppler relation are:
\begin{equation}
r[n]=\frac{\hat{R}_{\Sigma}^2[n]-D}{2\big(\hat{R}_{\Sigma}[n]+a[n]\big)},
\qquad
\hat{\vect{p}}_{\mathrm{TS}}[n]=\vect{p}_{\mathrm{R}}+r[n]\hat{\vect{u}}[n]
\label{eq:timestamp_position}
\end{equation}
\begin{equation}
\vect{v}^{\Transpose}[n]\vect{g}\big(\hat{\vect{p}}_{\mathrm{TS}}[n]\big)=-\lambda\hat{f}_{\mathrm{D}}[n]
\label{eq:velocity_projection}
\end{equation}
identifies only one velocity projection. For the scalar descriptor used with the independent estimates,
\begin{equation}
\hat{v}_{x,\mathrm{TS}}[n]=-
\frac{\lambda\hat{f}_{\mathrm{D}}[n]}{g_x\big(\hat{\vect{p}}_{\mathrm{TS}}[n]\big)},
\qquad g_x\big(\hat{\vect{p}}_{\mathrm{TS}}[n]\big)\neq0 
\label{eq:timestamp_velocity}
\end{equation}
Thus $\mathcal{T}_{\mathrm{TS}}=\{\hat{\vect{p}}_{\mathrm{TS}}[n],\hat{v}_{x,\mathrm{TS}}[n]\}_{n=1}^{N}$ forms the trajectory.
\vspace{3pt}
\subsection{EKF-Filtered Tracking}
\vspace{3pt}
The EKF tracks $\vect{x}[n]=[\vect{p}^{\Transpose}[n],\vect{v}^{\Transpose}[n]]^{\Transpose}\in\mathbb{R}^{6}$ using constant-velocity prediction \cite{ref_kalman_1960}. Over interval $\Delta t_n$, position advances by $\Delta t_n\vect{v}$ while velocity remains constant. Acceleration noise with covariance $\vect{Q}_a$ enters position/velocity through $\tfrac12\Delta t_n^2\vect{I}_3$ and $\Delta t_n\vect{I}_3$; $\vect{P}_{n|n-1}$ denotes predicted state covariance.
Let $\vect{d}_{\mathrm{T}}=\vect{p}-\vect{p}_{\mathrm{T}}$, $\vect{d}_{\mathrm{R}}=\vect{p}-\vect{p}_{\mathrm{R}}$, $r_{\mathrm{T}}=\norm{\vect{d}_{\mathrm{T}}}$, $r_{\mathrm{R}}=\norm{\vect{d}_{\mathrm{R}}}$, and $\rho=\sqrt{d_{\mathrm{R},x}^2+d_{\mathrm{R},y}^2}$. The observation model is:
\begin{equation}
\vect{z}[n]=\vect{h}\big(\vect{x}[n]\big)+\boldsymbol{\nu}[n],
\quad
\boldsymbol{\nu}[n]\sim\mathcal{N}(\vect{0},\vect{R}_n),
\label{eq:ekf_meas_model}
\end{equation}
with $\vect{R}_n=\diag(\sigma_R^2[n],\sigma_\phi^2[n],\sigma_\theta^2[n],\sigma_f^2[n])$ and nonlinear measurement function and the geometry vector can be defined:
\begin{equation}
\vect{h}(\vect{x})=
\begin{bmatrix}
r_{\mathrm{T}}+r_{\mathrm{R}}\\
\operatorname{atan2}(d_{\mathrm{R},y},d_{\mathrm{R},x})\\
\operatorname{atan2}(d_{\mathrm{R},z},\rho)\\
-\lambda^{-1}\vect{v}^{\Transpose}\vect{g}(\vect{p})
\end{bmatrix},
\quad
\vect{g}(\vect{p})=\frac{\vect{d}_{\mathrm{T}}}{r_{\mathrm{T}}}+\frac{\vect{d}_{\mathrm{R}}}{r_{\mathrm{R}}}
\label{eq:ekf_measurement_function}
\end{equation}
The Jacobian $\vect{H}_n=\partial\vect{h}/\partial\vect{x}$ is evaluated at $\hat{\vect{x}}_{n|n-1}$, with angular rows rescaled by $180/\pi$ for degree-valued angles. The standard EKF update corrects the predicted state by the gain-weighted innovation and updates its covariance in Joseph form.
Before the state update, the azimuth innovation is wrapped to its principal interval; unavailable measurement components are handled by removing the corresponding rows and columns from $\vect{z}[n]$, $\vect{h}$, $\vect{H}_n$, and $\vect{R}_n$. The filter is initialized from the first valid independent time-stamp reconstruction and then recursively regularizes position and velocity.

\section{Simulation Environment}
\label{sec:simulation_environment}

\begin{figure}[!ht]
    \centering
    \includegraphics[width=0.99\linewidth, height=0.8\linewidth]{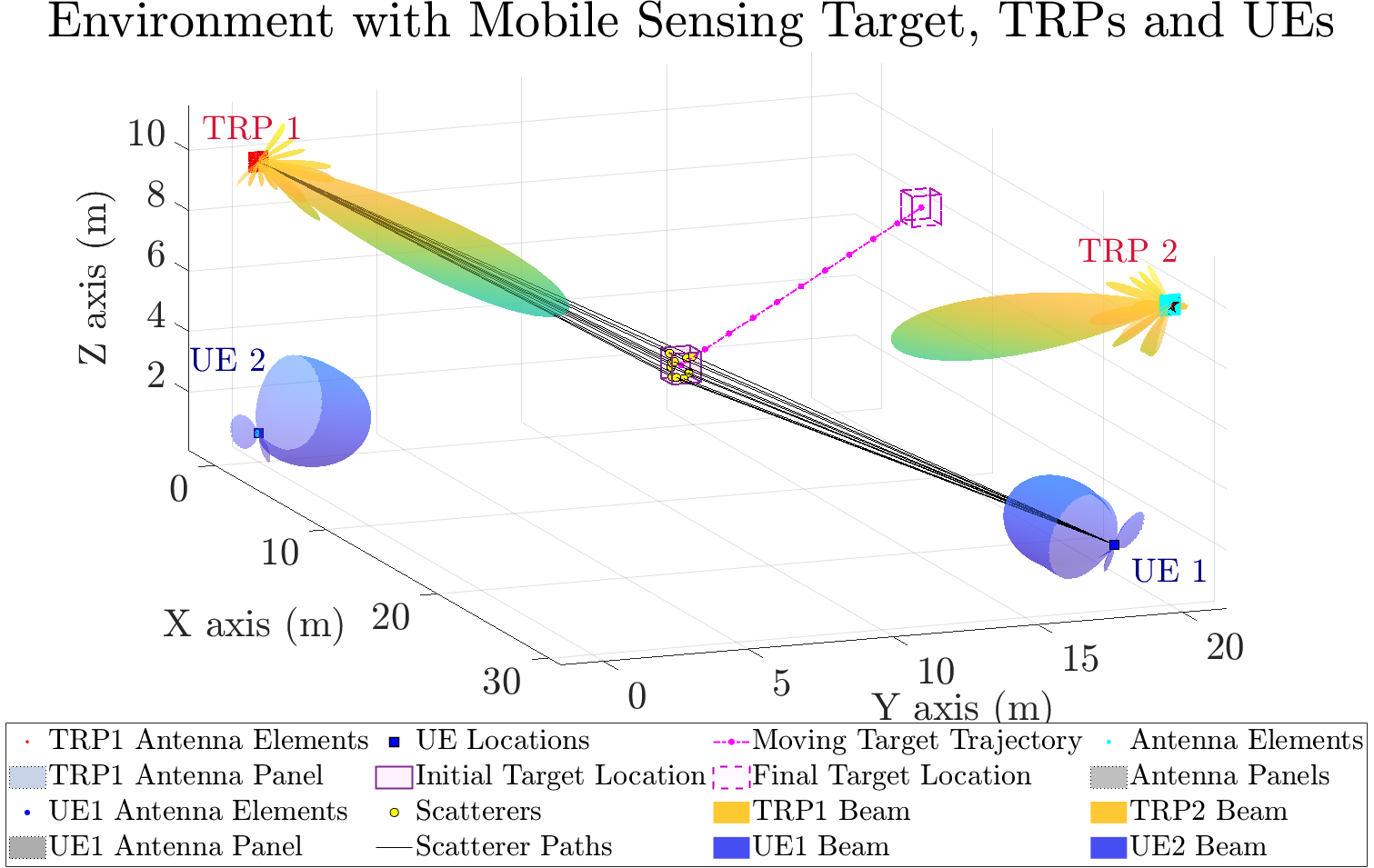}
    \caption{3D Simulation environment with TRP/UE locations, mobile ST trajectory, target scatterers, Tx--scatterer--Rx paths, and target-directed beams.}
    \label{fig:environment_view}
\end{figure}

Fig.~\ref{fig:environment_view} shows the three-dimensional ISAC environment used to evaluate the sensing modes in Fig.~\ref{SensingModes}. The scene contains two TRPs, two UEs, and a moving sensing target. For the fixed-node tracking example, the ST is modeled as a 1 m cubic extended target with $Q=20$ scatterer points. The same deployment is evaluated under five configurations: TRP monostatic sensing at TRP1, UE monostatic sensing at UE1, TRP--UE bistatic sensing from TRP1 to UE1, TRP--TRP bistatic sensing from TRP1 to TRP2, and UE--UE bistatic sensing from UE1 to UE2. In the bistatic cases, the first node denotes transmitter and the second node denotes receiver.

The fixed-node deployment follows Table~\ref{tab:simulation_environment_parameters}, with $8\times8$ TRP arrays and $2\times2$ UE arrays. For Fig.~\ref{fig:EKF_Tracking}, the deterministic target starts from $[6,10,5]^{\mathsf{T}}$ m and moves with velocity $[2,2,2]^{\mathsf{T}}$ m/s for 4 s at 0.3 s snapshot spacing. For Figs.~\ref{fig:CDF_Time_Stamp}--\ref{fig:CDF_Sensing_Modes}, initial target centers and velocities are sampled from Table~\ref{tab:simulation_environment_parameters}; trajectories leaving the environment box are rejected. A nominal SNR $s_0$ is drawn uniformly from [0:1:40] dB per trajectory and held fixed across its snapshots, modes, and beam choices. Each CDF pools all finite snapshot-wise 3D position errors across trajectories and sampled SNRs; it is not a fixed-SNR result.The beam-normalized estimator-input SNR is:
\begin{equation}
\begin{aligned}
\Gamma[n]&=10^{s_0/10}\max\!\left\{\frac{G[n]}{N_{\mathrm{T}}N_{\mathrm{R}}},10^{-3}\right\},\\
G[n]&=\left|\vect{w}_{\mathrm{T}}^{\mathsf{H}}\vect{a}_{\mathrm{T}}(\Omega_{\mathrm{T}}[n])\right|^2
\left|\vect{w}_{\mathrm{R}}^{\mathsf{H}}\vect{a}_{\mathrm{R}}(\Omega_{\mathrm{R}}[n])\right|^2,
\end{aligned}
\label{eq:mc_snr_definition}
\end{equation}
where $\Omega_{\mathrm{T}}[n]$ and $\Omega_{\mathrm{R}}[n]$ are target-center directions. Noise variance equals target-only power divided by $\Gamma[n]$, using element-averaged power before scalar combining for MUSIC and receive-combined slow-time power before the FFT for Doppler. Thus, $\Gamma[n]$ includes normalized Tx/Rx gains; it is not a pre-beamforming link-budget SNR.
Monte Carlo delay samples use the centroid in \eqref{eq:delay_centroid_discrete} plus zero-mean Gaussian noise with standard deviation
\begin{equation}
\sigma_{\tau}[n]=\max\!\left\{\frac{0.15}{B},\frac{1}{2\pi(B/\sqrt{12})\sqrt{2\Gamma[n]}}\right\}
\label{eq:mc_delay_std}
\end{equation}
Delays are bounded below by $\norm{\vect{p}_{\mathrm{T}}-\vect{p}_{\mathrm{R}}}/c+0.05\,\mathrm{ns}$. Delays are modeled as power-weighted centroids plus Gaussian noise. MUSIC uses 128 spatial snapshots and a $1^{\circ}$ angle grid; Doppler uses 128 slow-time samples at 10 kHz with a Hann window and a 2048-point FFT.
\begin{table}[!t]
\caption{Monte Carlo Simulation Configurations and Parameters}
\label{tab:simulation_environment_parameters}
\centering
\begin{tabular}{>{\raggedright\arraybackslash}p{0.42\linewidth} >{\raggedright\arraybackslash}p{0.48\linewidth}}
\toprule
Parameter & Value \\
\midrule
Carrier Frequency, BW, SCS & 35 GHz, 500 MHz, 240 kHz \\
TRP1 and TRP2 Position & $[0,0,10]^{\mathsf{T}}$ m, and $[30,20,10]^{\mathsf{T}}$ m\\
UE1 and UE2 Position & $[25,20,1]^{\mathsf{T}}$ m, and $[0,0,1]^{\mathsf{T}}$ m \\
ST Size and Num. of Scatterers & 1 m cubic, $Q=20$ \\
MC initial-center range & $[0,0,0]^{\mathsf{T}}$--$[25,20,20]^{\mathsf{T}}$ m \\
MC velocity range & $[-5,-5,-5]^{\mathsf{T}}$--$[5,5,5]^{\mathsf{T}}$ m/s \\
MC trials and snapshot times & $10^{6}$, 0:0.1:5 s, 51 snapshots \\
MC nominal SNR & $0{:}1{:}40$ dB; uniform per trajectory \\
\bottomrule

\end{tabular}
\vspace{-12pt}
\end{table}

\section{Simulation Results}
\label{sec:simulation_results}

The numerical results are organized in two parts. Fig.~\ref{fig:EKF_Tracking} corresponds to the deterministic fixed-node configuration described in Section~\ref{sec:simulation_environment}, whereas Figs.~\ref{fig:CDF_Time_Stamp}--\ref{fig:CDF_Sensing_Modes} correspond to the Monte Carlo simulations summarized in Table~\ref{tab:simulation_environment_parameters}.
\subsection{Fixed-Node Tracking Performance}

\begin{figure}[hb!]
    \centering
    \includegraphics[width=0.99\linewidth, height=0.8\linewidth]{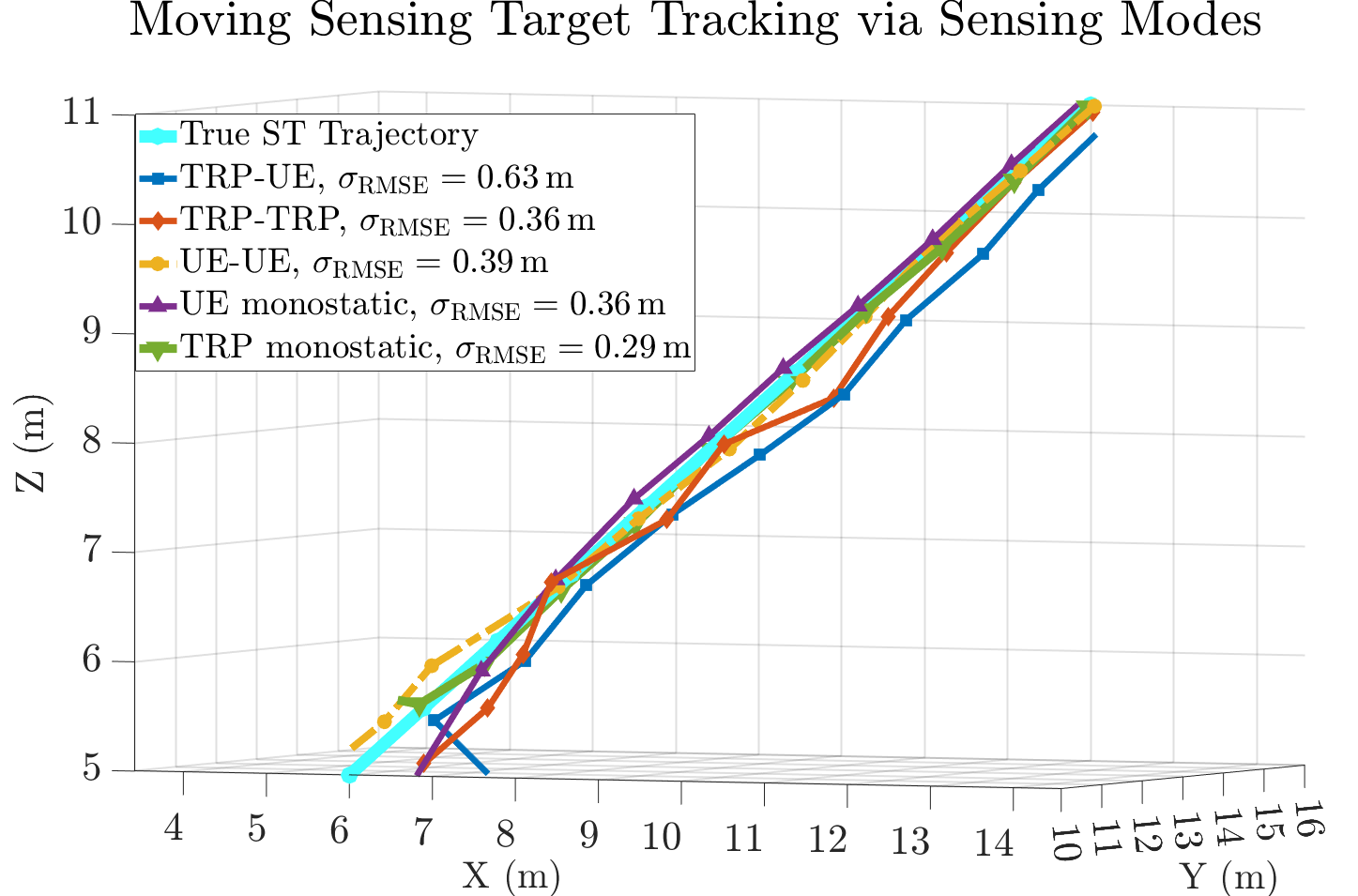}
    \caption{Fixed-node configuration: moving ST tracking performance via all sensing modes for the deterministic trajectory. The true ST trajectory is compared with EKF Tracks, and the legend reports RMSE for sensing modes.}
    \label{fig:EKF_Tracking}
\end{figure}

Fig.~\ref{fig:EKF_Tracking} compares the fixed-node tracking performance
across all sensing modes. All reconstructed trajectories closely follow the
true diagonal ST motion and achieve sub-meter accuracy. The TRP monostatic
mode gives the lowest error, with $\sigma_{\mathrm{RMSE}}=0.29$~m, followed
by TRP--TRP bistatic and UE monostatic modes with
$\sigma_{\mathrm{RMSE}}=0.36$~m, UE--UE bistatic with
$\sigma_{\mathrm{RMSE}}=0.39$~m, and TRP--UE with
$\sigma_{\mathrm{RMSE}}=0.63$~m.

\subsection{Monte Carlo Simulations}

\begin{figure}[!ht]
    \centering
    \includegraphics[width=0.99\linewidth, height=0.75\linewidth]{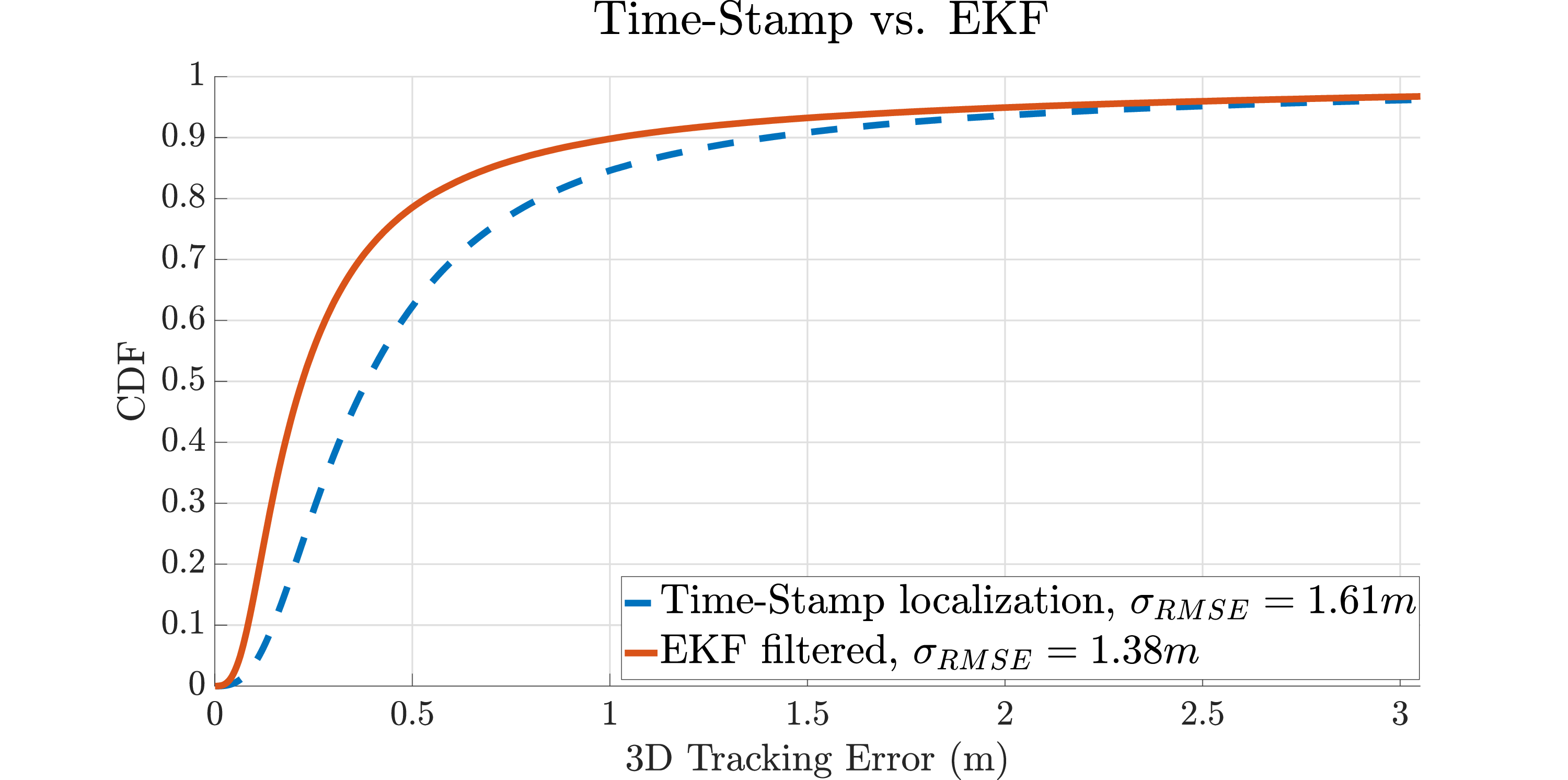}
    \caption{TRP--UE 3D position-error CDFs for time-stamp localization and EKF tracking with ST-directed beams; nominal SNRs of 0--40 dB are pooled.}
    \label{fig:CDF_Time_Stamp}
\end{figure}

The Monte Carlo CDFs quantify the statistical behavior over the randomized trajectories in Table~\ref{tab:simulation_environment_parameters}. Fig.~\ref{fig:CDF_Time_Stamp} compares independent time-stamp localization with EKF-filtered tracking. The EKF curve is shifted toward lower errors and yields $\sigma_{\mathrm{RMSE}}=1.38$ m, compared with $\sigma_{\mathrm{RMSE}}=1.61$ m for the independent time-stamp estimates. This corresponds to a relative RMSE reduction of approximately $14.3\%$, which confirms the benefit of using the constant-velocity motion model and recursive measurement fusion instead of processing each snapshot independently.

\begin{figure}[!ht]
    \centering
    \includegraphics[width=0.99\linewidth, height=0.75\linewidth]{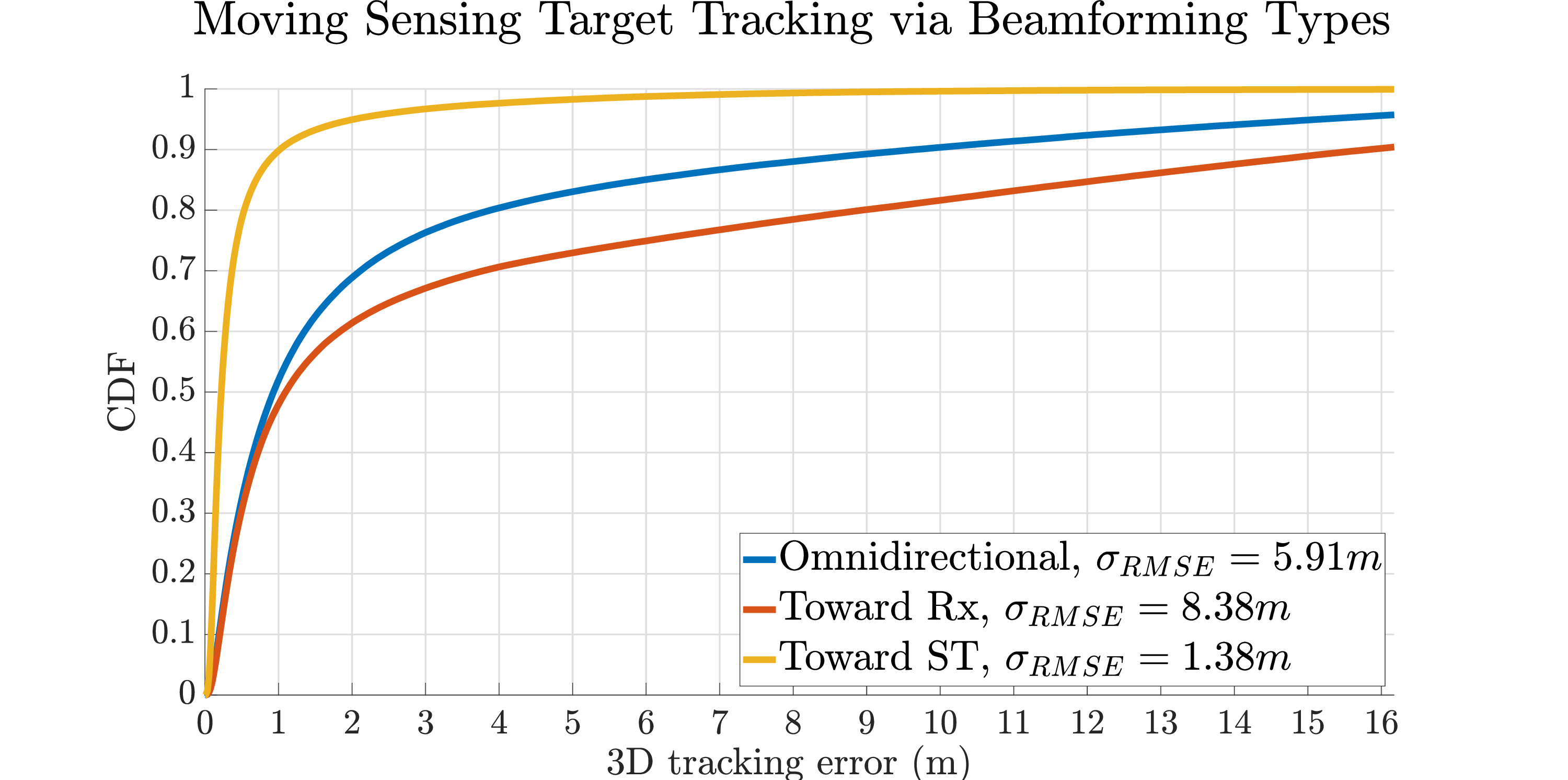}
    \caption{TRP--UE EKF position-error CDFs for omnidirectional, Rx-directed, and ST-directed beams; nominal SNRs of 0--40 dB are pooled.}
    \label{fig:CDF_Beamforming}
\end{figure}

Fig.~\ref{fig:CDF_Beamforming} compares the tracking performance for the three beamforming mechanisms. ST-directed beamforming achieves the steepest CDF and the lowest RMSE, $\sigma_{\mathrm{RMSE}}=1.38$ m. Omnidirectional operation gives $\sigma_{\mathrm{RMSE}}=5.91$ m, while Rx-directed beamforming gives $\sigma_{\mathrm{RMSE}}=8.38$ m. The result shows that directing the sensing beams toward the moving ST is essential for reliable echo extraction. Rx-directed beams can be unfavorable for passive target sensing because they may illuminate the communication receiver rather than the target reflection region, whereas omnidirectional operation avoids severe mispointing but sacrifices array gain.

\begin{figure}[!ht]
    \centering
    \includegraphics[width=0.99\linewidth, height=0.75\linewidth]{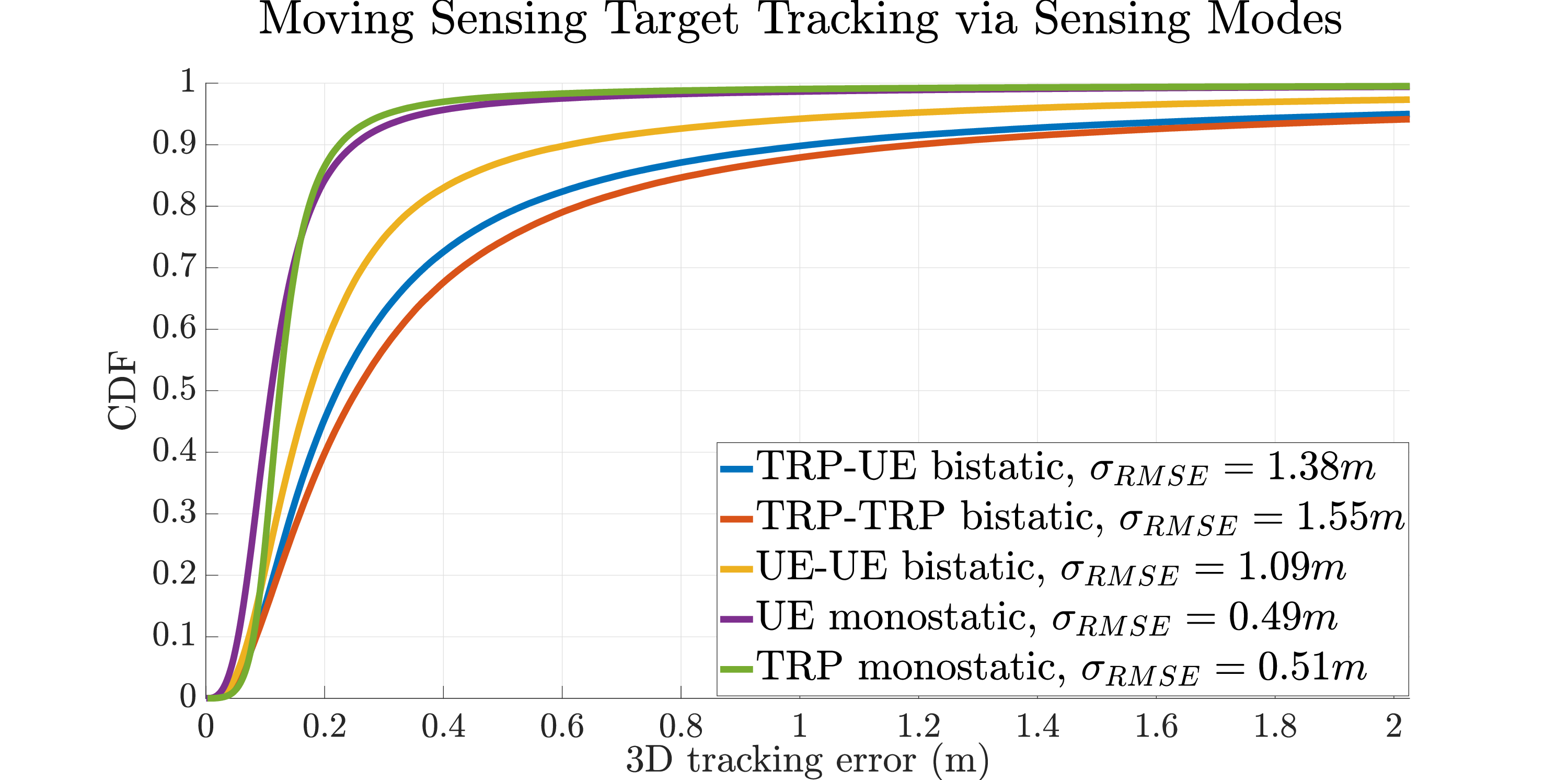}
    \caption{Sensing-mode EKF position-error CDFs with ST-directed beams}
    \label{fig:CDF_Sensing_Modes}
\end{figure}

Fig.~\ref{fig:CDF_Sensing_Modes} reports the Monte Carlo sensing-mode tracking performance with ST-directed beamforming and EKF tracking. The UE monostatic and TRP monostatic modes provide the best CDFs, with $\sigma_{\mathrm{RMSE}}=0.49$ m and $\sigma_{\mathrm{RMSE}}=0.51$ m, respectively. Among the bistatic modes, UE--UE bistatic sensing achieves $\sigma_{\mathrm{RMSE}}=1.09$ m, followed by TRP--UE bistatic sensing with $\sigma_{\mathrm{RMSE}}=1.38$ m and TRP--TRP bistatic sensing with $\sigma_{\mathrm{RMSE}}=1.55$ m. The stronger monostatic performance is consistent with the favorable co-located range-bearing geometry in the considered deployment, while the bistatic modes remain more sensitive to baseline orientation, target position, and the available Doppler projection. The similar monostatic RMSEs do not establish equal angular accuracy. Normalization in \eqref{eq:mc_snr_definition} removes the full coherent array-gain advantage from the prescribed SNR under ideal target alignment; aperture still affects MUSIC. At high effective SNR, both modes share the imposed delay-noise floor $0.15/B\approx0.300$ ns ($0.090$ m total-path-length standard deviation, not 3D tracking RMSE). Different node geometries, angle-grid quantization, scatterer centroids, and EKF filtering jointly influence error. Separate delay/angle/Doppler RMSEs would be required to identify the dominant limitation.

\section{Conclusion}
\label{sec:conclusion}

A beamformed PRS/OFDM ISAC framework was evaluated for moving extended-target sensing and tracking across multiple sensing modes. The sensing target was represented as a rigid cluster of scatterers, and the received echo was modeled as the coherent superposition of the corresponding Tx--scatterer--Rx paths. The measurement chain combined PRS matched filtering with delay centroiding, single-cluster 2D-MUSIC for azimuth/elevation estimation, FFT-based Doppler extraction, independent time-stamp localization, and EKF-based temporal tracking.

The fixed-node simulation confirmed that the proposed processing chain can reconstruct the moving ST trajectory in all considered sensing modes, with sub-meter RMSE for the deterministic trajectory. The Monte Carlo analysis showed three main trends. First, EKF filtering improves the tracking distribution relative to independent time-stamp localization, reducing the aggregate RMSE from $1.61$ m to $1.38$ m. Second, ST-directed beamforming is critical for robust sensing, reducing the RMSE to $1.38$ m compared with $5.91$ m for omnidirectional operation and $8.38$ m for Rx-directed beamforming. Third, for the considered geometry, monostatic UE and TRP sensing provide the best statistical tracking performance, with RMSE values of $0.49$ m and $0.51$ m, respectively, while the bistatic modes are more geometry-dependent.

Overall, the results demonstrate that active target-directed illumination, accurate delay-angle-Doppler extraction, and temporal filtering are key enablers for mobile ST tracking in ISAC networks. Future work will extend the evaluation to cluttered and non-line-of-sight environments, multiple simultaneous targets, adaptive beam management, and hardware-compliant NR waveform measurements.

\section*{Acknowledgment}
Funding was provided in part by the MSCA INTEGRATE and SNS JU 6G-DISAC Horizon Europe projects under Grant Agreement numbers 101072924 and 101139130, respectively, and by UK Research and Innovation (UKRI).


\begin{thebibliography}{99}

\bibitem{ref_liu_tcom_2020}
F. Liu, C. Masouros, A. P. Petropulu, H. Griffiths, and L. Hanzo, ``Joint Radar and Communication Design: Applications, State-of-the-Art, and the Road Ahead,'' \emph{IEEE Transactions on Communications}, vol. 68, no. 6, pp. 3834--3862, Jun. 2020, doi: 10.1109/TCOMM.2020.2973976.

\bibitem{ref_liu_jsac_2022}
F. Liu, Y. Cui, C. Masouros, J. Xu, T. X. Han, Y. C. Eldar, and S. Buzzi, ``Integrated Sensing and Communications: Toward Dual-Functional Wireless Networks for 6G and Beyond,'' \emph{IEEE Journal on Selected Areas in Communications}, vol. 40, no. 6, pp. 1728--1767, Jun. 2022, doi: 10.1109/JSAC.2022.3156632.

\bibitem{ref_zhang_survey_2022}
J. A. Zhang, M. L. Rahman, K. Wu, X. Huang, Y. J. Guo, S. Chen, and J. Yuan, ``Enabling Joint Communication and Radar Sensing in Mobile Networks - A Survey,'' \emph{IEEE Communications Surveys \& Tutorials}, vol. 24, no. 1, pp. 306--345, 2022, doi: 10.1109/COMST.2021.3122519.

\bibitem{ref_erkek_ojcoms_2026}
N. K. Erkek, M. Di Renzo, A. Shojaeifard, Y. Mestrah, R. Koirala, M. Heggo, and K. Shah, 
``TRP--UE Bistatic Sensing for 6G: Performance Analysis and Insights,'' manuscript submitted to \emph{IEEE Open Journal of the Communications Society}, under review.

\bibitem{ref_erkek_vtc2026_hybrid}
N. K. Erkek, M. Di Renzo, A. Shojaeifard, Y. Mestrah, R. Koirala, M. Heggo, and K. Shah, 
``Hybrid TRP--UE Sensing for Enhanced Target Localization,'' presented at the \emph{2026 IEEE 103rd Vehicular Technology Conference (VTC2026-Spring)}, Nice, France, Jun. 2026. Preprint: arXiv:2606.20413.

\bibitem{ref_cha_rel18}
H.-S. Cha, G. Lee, A. Ghosh, M. Baker, S. Kelley, and J. Hofmann, ``5G NR Positioning Enhancements in 3GPP Release-18,'' \emph{IEEE Communications Standards Magazine}, vol. 9, no. 1, pp. 22--27, Mar. 2025, doi: 10.1109/MCOMSTD.0001.2400006.

\bibitem{ref_italiano_tutorial}
L. Italiano, B. Camajori Tedeschini, M. Brambilla, H. Huang, M. Nicoli, and H. Wymeersch, ``A Tutorial on 5G Positioning,'' \emph{IEEE Communications Surveys \& Tutorials}, 2024, doi: 10.1109/COMST.2024.3449031.

\bibitem{ref_shahmansoori_twc_2018}
A. Shahmansoori, G. E. Garcia, G. Destino, G. Seco-Granados, and H. Wymeersch, ``Position and Orientation Estimation Through Millimeter-Wave MIMO in 5G Systems,'' \emph{IEEE Transactions on Wireless Communications}, vol. 17, no. 3, pp. 1822--1835, Mar. 2018.

\bibitem{ref_wymeersch_globecom_2018}
H. Wymeersch, N. Garcia, H. Kim, G. Seco-Granados, S. Kim, F. Wen, and M. Frohle, ``5G mmWave Downlink Vehicular Positioning,'' in \emph{Proc. IEEE GLOBECOM}, 2018, doi: 10.1109/GLOCOM.2018.8647275.

\bibitem{ref_wang_extended_target}
Y. Wang, M. Tao, and S. Sun, ``Beamforming Design for Integrated Sensing and Communication with Extended Target,'' in \emph{Proc. IEEE Globecom Workshops (GC Wkshps)}, Kuala Lumpur, Malaysia, Dec. 2023, pp. 1392--1397, doi: 10.1109/GCWkshps58843.2023.10465036.

\bibitem{ref_wang_3d_extended_target}
Y. Wang, M. Tao, and S. Sun, ``Cram\'er-Rao Bound Analysis and Beamforming Design for Integrated Sensing and Communication With Extended Targets,'' \emph{IEEE Transactions on Wireless Communications}, vol. 23, no. 11, pp. 15987--16000, Nov. 2024, doi: 10.1109/TWC.2024.3435864.

\bibitem{ref_luo_moving_target}
D. Luo, H. Wu, H. Luo, B. Lin, and F. Gao, ``Moving Target Sensing for ISAC Systems in Clutter Environment,'' in \emph{Proc. IEEE Wireless Communications and Networking Conference (WCNC)}, Dubai, United Arab Emirates, Apr. 2024, pp. 1--6, doi: 10.1109/WCNC57260.2024.10570670.

\bibitem{ref_pedraza_extended_tracking}
F. Pedraza, S. K. Dehkordi, J. Hauffen, S. Li, P. Jung, and G. Caire, ``Extended Target Parameter Estimation and Tracking with an HDA Setup for ISAC Applications,'' in \emph{Proc. IEEE Globecom Workshops (GC Wkshps)}, Kuala Lumpur, Malaysia, Dec. 2023, pp. 117--122, doi: 10.1109/GCWkshps58843.2023.10464687.

\bibitem{ref_schmidt_music}
R. O. Schmidt, ``Multiple Emitter Location and Signal Parameter Estimation,'' \emph{IEEE Transactions on Antennas and Propagation}, vol. 34, no. 3, pp. 276--280, Mar. 1986, doi: 10.1109/TAP.1986.1143830.

\bibitem{ref_kalman_1960}
R. E. Kalman, ``A New Approach to Linear Filtering and Prediction Problems,'' \emph{Journal of Basic Engineering}, vol. 82, no. 1, pp. 35--45, Mar. 1960, doi: 10.1115/1.3662552.

\end{thebibliography}
\end{document}